\documentclass{PoS}

\usepackage{lineno}

\usepackage{graphicx}
\usepackage{caption}
\usepackage{subcaption}
\usepackage{amsmath,amsfonts,amssymb}
\usepackage{nicefrac}
\usepackage{graphicx,wrapfig,lipsum}
\usepackage{lineno}
\title{ First search for GeV neutrinos from bright gamma-ray solar flares using the IceCube Neutrino Observatory }

\ShortTitle{First search for GeV neutrinos from bright gamma-ray solar flares using IceCube}

\author{
The IceCube Collaboration$^{\dagger}$\\
{$^{\dagger}$ \itshape \href{http://icecube.wisc.edu/collaboration/authors/icrc19_icecube}{http://icecube.wisc.edu/collaboration/authors/icrc19\_icecube}}\\
E-mail: \email{gdewasse@icecube.wisc.edu}
}

\abstract{
In response to a reported increase in the total neutrino flux in the Homestake experiment in coincidence with solar flares at the end of the eighties, solar neutrino detectors have searched for solar flare signals. Solar flares convert magnetic energy into thermal energy of plasma and kinetic energy of charged particles such as protons. As a consequence of magnetic reconnection, protons are injected downwards from the coronal acceleration region and can interact with dense plasma in the lower solar atmosphere, producing mesons that will subsequently decay into gamma rays and neutrinos at O(MeV-GeV) energies. 

The main motivation to search for solar flare neutrinos comes from their hadronic origin. As inherent products of high-energy proton collisions with the chromosphere, they are a direct probe of the proton accelerated towards the chromosphere. Using a multi-messenger approach, it is therefore possible to constrain the proton acceleration taking place in the solar flares, including the spectral index of the accelerated flux and its shape.

We present the results of the first search for GeV neutrinos emitted during solar flares carried out with the IceCube Neutrino Observatory. We present a new approach which allows us to strongly lower the energy threshold of IceCube, originally designed to detect 10 GeV - PeV neutrinos. We compare the results with theoretical estimates of the corresponding flux. 
\\

\vspace{4mm}
{\bfseries Corresponding authors:}
\speaker{Gwenhael de Wasseige}$^{1}$\\
{$^{1}$ \itshape Laboratoire APC, Paris-Diderot, France}

}

\FullConference{36th International Cosmic Ray Conference -ICRC2019-\\
		July 24th - August 1st, 2019\\
		Madison, WI, U.S.A.}

\begin{document}
\section{Multi-Messenger Signal Emitted by Solar Flares}
While multi-messenger astronomy has recently recorded major breakthroughs (e.g., ~\cite{TXS}), the Sun has been detected through several messengers for decades.
Both the quiescent and the active Sun are studied through the electromagnetic radiation they emit across the entire spectrum. 
Special attention will be given to the gamma rays produced by high-energy protons, accelerated in solar flares, when they collide with dense layers of the solar atmosphere. These gamma rays can be detected by the Fermi-LAT satellite, which has contributed to significantly increase the fraction of solar flares detected in the high-energy range~\cite{fermi-lat-10-years}.
Protons, electrons, and neutrons have also been detected during energetic phenomena linked with solar activity, such as solar flares. Solar neutrinos, however, have only been detected in the MeV range so far, being produced by fusion reactions happening in the core of the Sun. Despite several searches perfomed in neutrino telescopes since the late eighties~\cite{homestake, bahcall, kamiokande, sno}, no significant signal from solar flares has been confirmed so far. 

This work will describe the interest of searching for GeV neutrinos produced in solar flares. We present an optimized choice of solar flares to be studied, based on the tight link existing between gamma rays and neutrinos. The last part of these proceedings presents the first results of a solar flare neutrino search carried out with the IceCube Neutrino Observatory~\cite{jinst}. 
We conclude with the perspectives for the next solar cycle. 
\section{The Potential Physical Constraints Brought by Neutrinos}
Solar flares convert magnetic energy into plasma heating and kinetic energy of charged particles such as protons~\cite{hudson}. As a consequence of the magnetic reconnection, protons are injected downwards from the coronal acceleration region and can interact with the dense plasma in the lower solar atmosphere. The consequent processes are indicated in Eq.~\ref{reaction} where the energy threshold is 280 MeV for p-p and p-n.

\begin{equation}
p\, +\,  p \,{~\rm or~} \, p\, + \, \alpha
\longrightarrow 
\left\{
\begin{array}{l}
 \pi^+ + X; \\
  \pi^0 \,+ X; \\
    \pi^- + X; \\
    \end{array}
\right.\\
\label{reaction}
\\
\begin{array}{l}
\pi^+ \longrightarrow \mu^+ + \nu_{\mu} ;~ \mu^+ \longrightarrow e^+ + \nu_e + \bar\nu_{\mu} \\
\pi^0 \longrightarrow 2 \gamma \\
\pi^- \longrightarrow \mu^- + \bar{\nu}_{\mu}; ~\mu^- \longrightarrow e^- + \bar{\nu}_e + \nu_{\mu} \\
\end{array}
\end{equation}

\subsection{A specific solar flare selection for neutrino searches}
Following Eq.~\ref{reaction}, we see that pion production generates both neutrino and
gamma-ray emissions. We want to focus only on solar flares that emit neutrinos and thus use the gamma-ray observations to pick the most relevant candidates for our neutrino search. 
Fortunately, pion-decay products seem to dominate the gamma-ray spectrum above 100 MeV~\cite{vilmer}. Furthermore, Fermi-LAT has the ability to detect gamma rays above 100 MeV coming from the Sun. A spectral analysis confirms that the major contribution of the LAT observations was indeed consistent with pion decay emissions~\cite{ackermann}. 

We assume that neutrino emission is coincident with significant pion decay signals detected in Fermi-LAT during solar flares. This yields a significant difference in the studied sample compared to previous solar flare selections~\cite{kamiokande, sno} based on the X-ray
flux, considering that Fermi-LAT detects in average only 5\% of the M- and X-solar flares
detected by GOES~\footnote{This number has been obtained averaging the number of solar flares detected by Fermi-LAT between 2011 and 2015 and the number of M and X-class flares in GOES during the same period. It has to be noted that, since the Sun is not always in the field of view of Fermi-LAT, this number does not directly convert to the fraction of pion flares.}.

Fermi-LAT has also been able to detect an impulsive phase in some of the latest solar flare events. The light curve usually shows a short high-intensity peak on top of a longer period with a lower flux.  The peak is labelled the impulsive phase. Moreover, the analysis of the gamma rays detected during these short phases reveals a relatively hard initial proton spectrum, with a spectral index around 3, as well as an enhanced gamma ray yield~\cite{fermilde}.
In contrast, the long duration emissions manifest a softer proton spectral index (typically between 4 and 6) and a spread of the gamma-ray emission over several hours.
Focusing on the impulsive phase of bright events of the last solar cycle therefore increases the chance of a neutrino detection in coincidence with solar flares.

\begin{figure}[t!h!]
    \centering
    \includegraphics[width=0.55\textwidth]{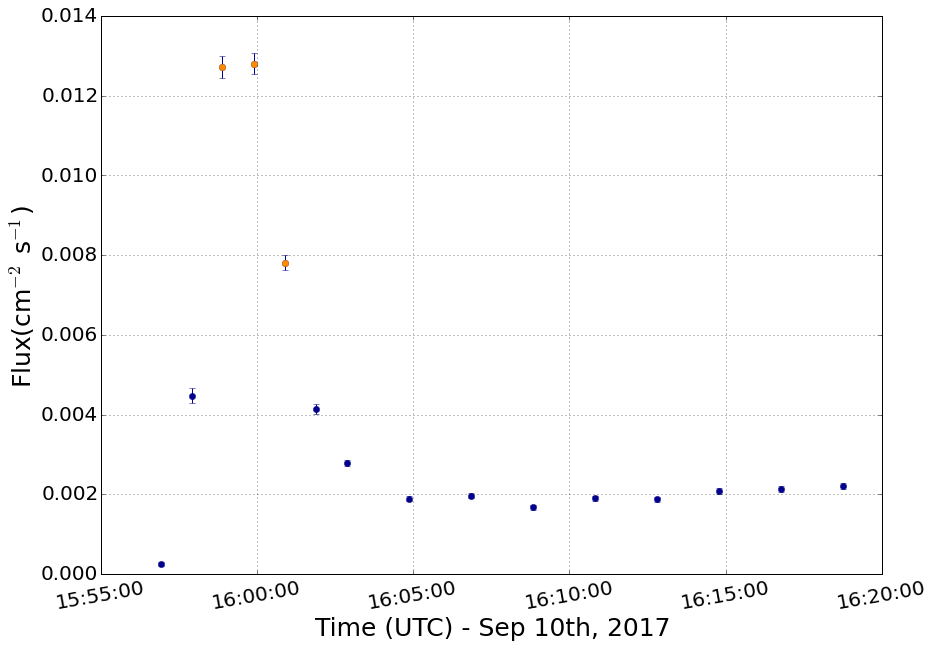}
    \caption{ Selected time window (orange points) compared to the initial gamma-ray light curve of the impulsive phase of the solar flare of Sept 10th, 2017. Data are provided by the Fermi-LAT collaboration.\label{lightcurve}}
\end{figure}

These criteria resulted in a list of 6 promising candidates for our first neutrino search. The details of each analyzed solar flare are reported, are reported in Table~\ref{table-sf}. The choice of the time window and duration of each solar flare was made in view of maximizing the signal-to-noise ratio in IceCube and is illustrated for the solar flare of Sept 10th, 2017 in Fig.~\ref{lightcurve}.

\begin{table}
\centering
\caption{Optimized time window for neutrino searches. For Sept 6th 2017, our optimization method led to consider the long-duration emission in this solar flare.}
\begin{tabular}{c|c| c |c}
Date & Selected time window & Duration (minutes) & Fraction observed\\
\hline
March 7th, 2012  & 00:41:22 - 01:21:22 & 40 & 85\%\\
February 25th, 2014  & 01:07:30 - 01:32:30& 25 & 97\%\\
September 1st, 2014 & 11:07:00 - 11:21:00&  14 &95.5\%\\
September 6th, 2017 & 13:23:03  - 22:00:37 & 515&87\%\\
September 10th, 2017 & 15:58:54 - 16:02:52& 5.96 &45\%\\
\end{tabular}
\label{table-sf}
\end{table}

\subsection{Added value of neutrino search in solar flare physics}
The main interest in searching for solar flare neutrinos comes from their hadronic origin: being inherent products of high-energy proton collisions in the chromosphere, they represent a direct probe of the proton acceleration. Several studies (see e.g. \cite{moriondproceeding, thesis, fargion}) have demonstrated that this neutrino flux could extend from MeV up to a few GeV. Focusing on the upper part of the solar flare neutrino spectrum would allow us to probe the proton acceleration up to the highest energies that can be reached within the solar flare environment.
In the following, we will assume the accelerated proton flux can be modeled using the following form $\frac{d\phi}{dE} = A E^{-\delta} H(E_{max} - E)$, where A is a normalization constant, $\delta$ represents the spectral index, and E$_{max}$ is the upper cutoff in a Heaviside function. 

\begin{figure}[t!]
    \centering
    \includegraphics[width=0.55\textwidth]{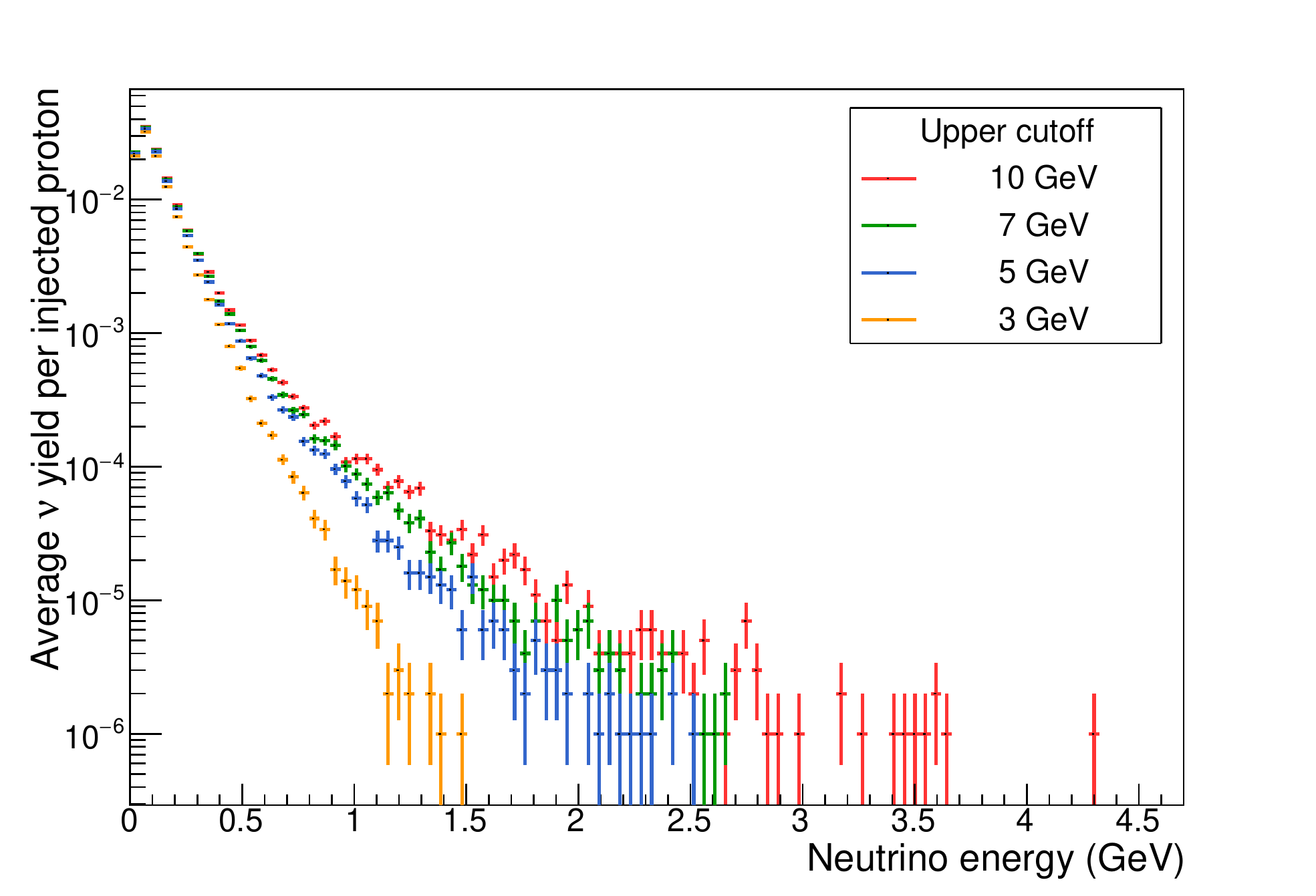}
    \caption{ Energy distribution for neutrinos depending on the upper cutoff in the accelerated proton spectrum with a spectral index of 3.\label{uppercutoff-spectrum}}
\end{figure}

The proton spectral index has been extracted from gamma-ray observations by Fermi-LAT for different phases of the solar flare~\cite{fermilde}, but there are so far no constraints on the value of the upper cutoff. The effect of this upper cutoff on the subsequent neutrino flux is illustrated in Fig.~\ref{uppercutoff-spectrum}, where the color points show the average neutrino yield per injected proton with $\delta$ = 3  and different realistic values of $E_{max}$. A higher cutoff value leads to a higher neutrino yield in the GeV energy range and would thus lead to a larger signal in sensitive neutrino telescopes.
Coupling Fermi-LAT and IceCube observations, one has therefore the potential to  constrain both this upper cutoff and the spectral index by fitting the gamma-ray spectrum and using the strength of the neutrino signal detected during this solar flare.

\section{The IceCube Neutrino Observatory: from $\geq$ TeV to GeV}\label{icecube}
The IceCube Neutrino Observatory is currently one of the best places to search for GeV neutrinos emitted during solar flares. IceCube is a cubic-kilometer neutrino detector installed in the ice of the geographic South Pole between depths of 1450~m and 2450~m~\cite{jinst}. It consists of the 5160 digital optical modules (DOMs) distributed along 86 strings sparsed across the detector volume.
A lower energy infill, the DeepCore subarray, includes 8 densely instrumented strings with smaller spacing between its optical modules (7~m versus 17~m in the IceCube strings) and its strings (72~m on average versus 125~m in IceCube)~\cite{deepcore}. When a neutrino interacts in the neighborhood of the detector, the subsequent electromagnetic and/or hadronic cascade emit Cherenkov photons that can be detected by one or several DOMs.

While IceCube was originally dedicated to observe TeV neutrinos, the collaboration has demonstrated the ability to extend the sensitivity to a larger energy range using DeepCore. Since the observation of the first astrophysical neutrinos in 2013~\cite{hese}, several noteworthy limits have been set on, among others, the existence of sterile neutrinos~\cite{sterile} and the spin-dependent WIMP-nucleon cross section~\cite{solar} as well as competitive measurements of neutrino oscillation parameters~\cite{thetaparameter}.  The collaboration has also joined the worldwide multimessenger effort , studying the highest energetic events in our Universe~\cite{amon, realtime} and leading to the first joint gamma-ray - neutrino observation~\cite{TXS}.


\subsection{Selection of GeV events}~\label{lowen}
As previously mentioned, a subarray of IceCube, DeepCore, enables the study of low energy neutrino interactions. Besides a higher density of optical modules, a softer trigger condition has been implemented in DeepCore in view of increasing the sensitivity to lower neutrino energies. We have simulated the interactions of GeV neutrinos in IceCube using the interaction physics in GENIE 2.8.6~\cite{genie1}, which includes the nuclear model, cross sections, and hadronization process~\cite{genie2} based on KNO~\cite{genie3} and PYTHIA~\cite{genie4}. 
A detailed description of the event selection developed in view of extracting GeV neutrino events from the IceCube data is presented in ~\cite{icrc-gw}. The main ingredients used in this selection are the amount of detected photoelectrons from the neutrino interaction and the presence or absence of causality between the DOMs that have detected the event.

\begin{figure}[t!]
    \centering
    \includegraphics[width=0.55\textwidth]{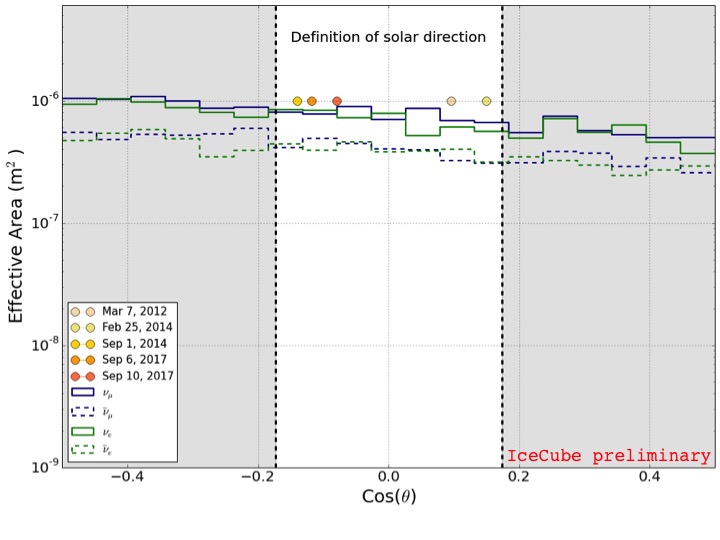}
    \caption{  Effective area for $\nu_e$ and $\nu_{\mu}$ at the final level of the event selection. The plot shows that the effective area is constant in the zenith band considered in the analysis.\label{effectivearea}}
\end{figure}

The selection allows to reduce the data rate down to 20~mHz, representing a reduction of 5 orders of magnitude, while keeping more than 40\% of the GeV neutrino events. The final rate is slightly larger than the expectation from atmospheric neutrinos, estimated to occur at the mHz level. The selection is sensitive both to single transient events and to an event-stacking analysis. The IceCube effective area for neutrinos passing the described selection is shown in Figure~\ref{effectivearea}.

\section{Results of the first GeV neutrino solar flare search performed in IceCube} It is  currently not feasible to reconstruct the incoming direction of the neutrino events passing the above selection. We thus develop a new approach, whose the general idea is to monitor the rate of GeV-like events in IceCube and to search for an increase in this rate during a solar flare.
In order to evaluate the background level and its natural fluctuations, we integrate 8 hours of data prior to the solar flare of interest. This background window yields a 13\% uncertainty on the 0.02 Hz background rate. We do not use the time region directly before the solar flare to
avoid including a potential precursor neutrino emission in the background
level determination.

In view of the rather large statistical uncertainties, we use the statistical test proposed by Li and Ma~\cite{liandma}. This method has been developed to estimate the significance of events in a certain time region, with a null-hypothesis being that no extra source exists.
The significance of the observed results is calculated using three parameters: the number of events \textit{N$_{\text{on}}$} in the solar flare region, \textit{N$_{\text{off}}$} for the number of events in the time region prior to the solar flare onset and $\alpha$ = $\frac{t_{\text{on}}}{t_{\text{off}}}$, where $t_{\text{off}}$ is 8~hours, and $t_{on}$, the selected time window during the solar flare.

\begin{figure}[t!]
    \centering
    \includegraphics[width=0.55\textwidth]{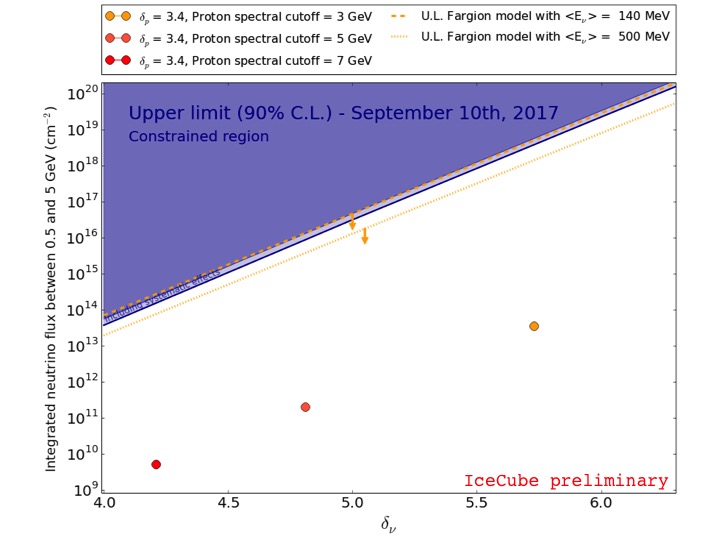}
    \caption{ Comparison of the experimental upper limit derived for March 7th, 2012 and the corresponding theoretical predictions. The x-axis represents the spectral index of the neutrino flux and the y-axis the normalisation constant of the neutrino flux. The orange points show the output of our simulation when assuming a proton spectral index of 3.2, derived from gamma-ray observations. The orange line shows the predictions from Fargion, with E$_{fl}$ =10$^{32}$ erg and $\langle$ E$_{\nu}$ $\rangle$=140 MeV (dashed) and 500 MeV (dotted).\label{UL}}
\end{figure}

\begin{table}[t]
\begin{center}
\caption{Number of off-source and on-source IceCube events as well as the corresponding significance obtained for each solar flare. For comparison, we add the expected amount of events (N$_{\text{on-expected}}$  ) based on the background rate of the analysis (i.e., 0.02~Hz) and the considered duration for each solar flare.}
\begin{tabular}{c| c | c| c | c }\hline
{\textbf{Event}} & {\textbf{N$_{\text{off}}$} } & {\textbf{N$_{\text{on}}$} } &{\textbf{N$_{\text{on-expected}}$} } & \textbf{Significance S} \\
\hline
Mar 7th, 2012 &761& 67 & 62&0.43 $\sigma$ \\
Feb 25th, 2014  &611 & 27& 32&0.86 $\sigma$  \\
Sep 1st, 2014 & 621 &21 & 18&0.65 $\sigma$ \\
Sep 6th, 2017 &  569& 639 &620& 0.79 $\sigma$ \\
Sep 10th, 2017 &  529& 5 & 6&0.64 $\sigma$ \\
  \hline
\end{tabular}
\label{tab:resultsnumberofevents}
\end{center}
 \end{table}
Table~\ref{tab:resultsnumberofevents} shows the number of off-source and on-source IceCube events as well as the corresponding significance $S$. 
None of the studied solar flares led to a significant signal when using the analysis presented in this work. We can therefore derive upper limit on the neutrino flux based on the potential number of signal events obtained using N$_{\text{on}}$ - $\alpha$ N$_{\text{off}}$ and the effective areas presented in Section~\ref{lowen}. 
The exact spectral index $\delta_{\nu}$ of the neutrino spectrum is not precisely known but could be estimated between 4 and 6 using a Geant4~\cite{geant4} simulation of proton-nucleus interaction in a solar environment\footnote{We assumed a mass fraction of 75\% of hydrogen and
25\% of helium for the solar atmosphere, which is a standard assumption for the solar
surface.} described in~\cite{thesis}. It is convenient to present the neutrino upper limit in the 
($\delta_{\nu}$, C) parameter space, where C is the integrated neutrino flux between 0.5 and 5 GeV. The obtained upper limit corresponding to the solar flare of Sept 10th, 2017 is shown in Figure~\ref{UL}. Similar results have been obtained for the other studied solar flares.

We can compare the values obtained with theoretical predictions~\cite{thesis, fargion}. 
Estimating the expected neutrino fluence from Eq. 2.11 in~\cite{fargion}, using $E_{fl}$ = 10$^{32}$ erg and $<E_{\nu_e}>$= 140~MeV as well as an optimistic $<E_{\nu_e}>$= 500~MeV, we obtain 410$\times$ 10$^3$ and 114$\times$ 10$^3$ neutrinos cm$^{-2}$ respectively, both of which are significantly larger than the estimate obtained in~\cite{thesis}.
The large difference in the two predictions is mainly due to the different assumption made on the fraction of the solar flare energy distributed to the protons accelerated up to the highest energies and a different fraction of proton energy given to the subsequent pions. 
For all the solar flares, except September 6th time window, which targets the long-duration emission of the solar flare and therefore cannot be compared with the models, the experimental upper limit constrains Fargion's prediction~\cite{fargion} when assuming an average energy of 140~MeV. The optimistic 500~MeV line is slightly below the reach of the current sensitivity. The other prediction~\cite{thesis} however stands far below the current reach of IceCube and cannot currently be tested. We note that the simulation developed in~\cite{thesis} successfully reproduces the integrated flux detected by Fermi-LAT for different proton spectral indices.

\section{Summary and Perspectives}

These proceedings report on the first search for a solar flare signal using the IceCube Neutrino Observatory. While no significant detection was made, we have demonstrated the potential interest of searching for solar flare neutrinos and validated the innovative approach required to allow IceCube to be sensitive in the GeV energy range. 

The large neutrino telescope landscape is expected to change in the coming years, with among others, the deployment of the IceCube-Upgrade~\cite{icrc-upgrade} within IceCube. This detector will demonstrate a lower detection threshold together with enhanced reconstruction capabilities because of the multi-photomultiplicator geometry of its sensors. We therefore expect the coming solar cycle to benefit from these additional sensors and the potential synergy with the work presented in these proceedings in view of reaching the required sensitivity to detect the solar flare neutrino flux.

\end{document}